\documentclass[aps,prl,twocolumn,showpacs,groupedaddress]{revtex4} 
\usepackage{graphicx}  
\usepackage{dcolumn}   
\usepackage{bm}        
\usepackage{amssymb}   

\begin{document}



\title{On the apparent lack of  power in the CMB anisotropy at large angular scales}
\author{Amir Hajian} 
\email{ahajian@princeton.edu}
\affiliation{                                                                
\centerline{ Department of Physics, Jadwin Hall, Princeton University,
    Princeton, NJ 08542,}    }  
\affiliation{                                                                
\centerline{ Department of Astrophysical Sciences, Peyton Hall,
   Princeton University,
    Princeton, NJ 08544.}    }  
\date{\today}

\begin{abstract}
We study the apparent lack of power on large angular scales in the WMAP data. We confirm that although there is no apparent lack of power at large angular scales for the \emph{full-sky} maps, the lowest multipoles of the WMAP data happen to have the magnitudes and orientations, with respect to the Galactic plane, that are needed to make the large scale power in \emph{cut-sky} maps surprisingly small. Our analysis shows that most of the large scale power of the observed CMB anisotropy maps comes from two regions around the Galactic plane ($\simeq 9\%$ of the sky). One of them is a cold spot within $\sim 40^\circ$ of the Galactic center and the other one is a hot spot in the vicinity of the Gum Nebula. If the current full-sky map is correct, there is no clear deficit of power at large angular scales and the alignment of the $l=2$ and $l=3$ multipoles remains the primary intriguing feature in the full-sky maps. If the full-sky map is incorrect and a cut is required, then the apparent lack of power remains mysterious. 
 Future missions such as \emph{Planck}, with a wider frequency range and greater sensitivity, will permit a better modeling of the Galaxy and will shed further light on this issue. 

\end{abstract}

\maketitle
\begin{figure}
 \includegraphics[width=0.5\textwidth]{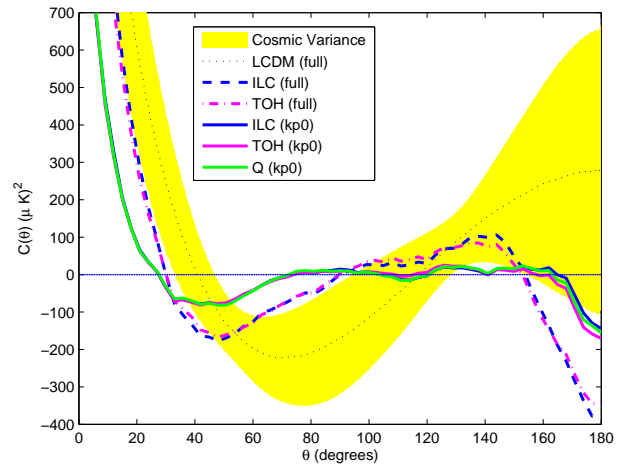}
 \caption{\label{fig:Ctheta} The two-point correlation functions for WMAP data. Cut-sky maps have surprisingly low power on angular scales larger than $60^\circ$.  This is consistently seen in the ILC map, TOH map and WMAP foreground reduced maps for Q, V and W frequency bands masked with kp0 intensity mask. Two-point correlation function for full-sky ILC and TOH maps are also shown. Unlike masked maps, they are not flat on large angular scales.}
 \end{figure}

 
Over the last few years, detailed measurements of the anisotropies in the Cosmic Microwave Background radiation (CMB) have provided us a wealth of information. Recently the \emph{Wilkinson Microwave Anisotropy Probe} (WMAP)  \cite{Hinshaw:2006ia,Jarosik:2006ib,Page:2006hz,Spergel:2006hy} mission released maps of 3 years of data of the full sky in five different frequency bands. While the success of the standard inflationary cosmology in explaining these observations combined with other astronomical data is remarkable, there are apparent discrepancies between prediction and observations on large angular scales. 
One of these discrepancies is the lack of correlated signal on large angular scales in cut-sky maps of WMAP data. It seems that masking the sky with a Galactic mask reduces the correlation on large scales. This feature  was also clearly seen by COBE \cite{Hinshaw:1996ut}.  Figure \ref{fig:Ctheta} shows the measured two-point correlation in full- and cut-sky maps together with the prediction of the best fit LCDM model. The flatness of the two-point correlation of the WMAP data on large angular scales was first quantified by the \emph{a posteriori} statistic of \cite{Spergel:2003cb} for the first-year data and in more detail by \cite{Copi:2006tu} for the 3-year data. It was reported that the  large-angle correlations, for $\theta>60^\circ$, are unusually weak at the $99.85\%$ confidence level and it is more siginificant in three year data than in the first-year data \cite{Copi:2006tu}. 

The two-point correlation function, $C(\theta)$, of a CMB anisotropy map is the Legendre transform of its power spectrum $C_l$,
\begin{equation}
  \label{eq:4}
  C(\theta) = \frac{1}{4\pi}\sum_{l=2}^{\infty}(2l+1) C_l P_l(\cos\theta).
\end{equation}
 On large angular scales,  $C(\theta)$ is sensitive to the low order multipoles and  low-$l$ features in the angular power spectrum are projected onto large-angle patterns in the two-point correlation.  But the disadvantage of using $C(\theta)$ is that it is highly correlated at different angular scales. Therefore a covariance analysis is suggested for reliable statistical studies of the two-point correlation \cite{Ganga:1994ui, Bernui:2006ft}. 

To quantify the lack of power on large scales, \cite{Spergel:2003cb} suggested the sum of the square of the two-point correlation  $\int_{-1}^{1/2} C^2(\theta)d\cos\theta$. That is a simple and useful statistic but since $C(\theta)$ is correlated on different angular scales, it is better to use a covariance weighted sum of the square of the two-point correlations defined by the following four point statistic
\begin{equation}
  \label{eq:1}
  A(x) = \int_{-1}^{x}\int_{-1}^{x} C(\theta) F^{-1}(\theta,\theta')  C(\theta') d\cos{\theta} d\cos{\theta'},
\end{equation}
in which $ F(\theta,\theta') $ is the covariance between different bins in 
$C(\theta)$ defined as
\begin{equation}
  \label{eq:2}
   F(\theta,\theta') = \langle \tilde{C}(\theta) \tilde{C}(\theta') \rangle.
\end{equation}
where $ \langle \cdots \rangle$ represents the ensemble average and $\tilde{C}(\theta) = C(\theta) -  \langle C(\theta) \rangle$.
Note that $A(1/2)$ in the limit of uncorrelated $C(\theta)$, will be the same as the $S_{1/2}$-statistic originally proposed by \cite{Spergel:2003cb}.

 We use the following maps and masks for our analysis:
\begin{itemize}
\item The full-sky WMAP Internal Linear Combination (ILC) map released by the WMAP team.
\item The three-year foreground-cleaned map of Tegmark et al. \cite{Tegmark:2003ve} (TOH map) \footnote{This map can be found here: http://space.mit.edu/home/tegmark/wmap.html}.
\item WMAP foreground reduced maps for Q, V and W frequency bands, co-added from each differencing assembly using noise weighting (see \cite{Hinshaw:2006ia}).
\item Kp0 intensity mask. This mask excludes the Galactic region (in which the K-band intensity is high) and also $0.6^{\circ}$ around known point sources ($\sim$23.46\% of the pixels from the maps) at r9 resolution corresponding to HEALPix resolution $N_{side}=512$. WMAP maps and the mask are available on LAMBDA \footnote{http://lambda.gsfc.nasa.gov/} as part of the three year data release.
\end{itemize}
\begin{figure}
 \includegraphics[width=0.5\textwidth, angle=0]{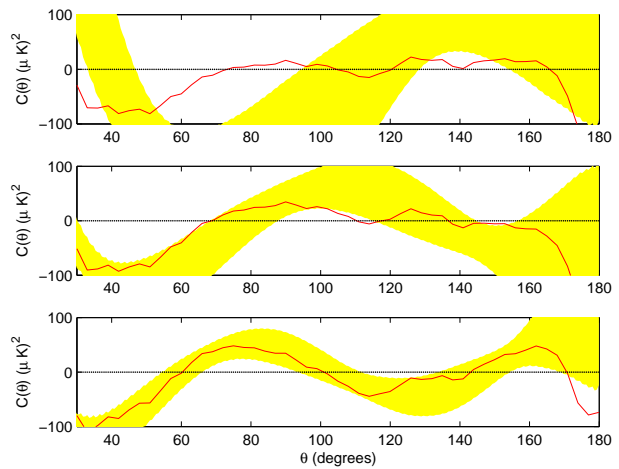}
 \caption{\label{fig:multipoles}  Apparent lack of power on large angular scales in cut-sky ILC (\emph{top}) may be more than just a low quadrupole. Removing the quadrupole does not change the shape of the two-point correlation much (\emph{middle}). But removing $l=2,3$ together has a bigger effect (\emph{bottom}). }
 \end{figure}
All maps and masks are at r9 resolution from the three year data of WMAP. 
The two-point correlations are computed using eqn.\ref{eq:4} and the covariance matrix $F(\theta,\theta')$, is computed from simulations of the best fit LCDM model for each mask.
Computing $A(x)$ for the above maps shows that $A(x)$ for cut-sky maps is much smaller than for full-sky maps. This is expected because of the small values of  $C(\theta)$ on large angular scales in masked maps.

To assign a statistical significance to the smallness of $A(x)$ computed from the data, we perform comparisons against 10,000 random simulations of statistically isotropic and Gaussian CMB anisotropy maps based on the best fit LCDM model to the WMAP data. Since $A(x)$ is positive definite, the statistical significance of $A(x)$ at each point is defined as the fraction of the simulations that have a smaller $A(x)$ than the data, $P(A_{sim}(x)<A_{data}(x))$. The results are shown in Table \ref{tab:table1} in which we have made an \emph{a posteriori} choice for $x$ to be $x\simeq 0.53 \, (\theta=58^\circ)$. This is the point at which $A(x)$ of the masked ILC map reaches its minimum. We will use $A(1/2)$ as a short name for $A(x=0.53)$.  

The full-sky maps are not anomalous; $8\%$ of 10,000 simulations have an $A(x)$ smaller than that of the full-sky WMAP maps (see \cite{Bernui:2006ft} for a different statistic but similar conclusion). The odd shape of the full-sky correlation, and in particular the anticorrelation at $\theta=\pi$, is due to the fact that even multipoles are weaker than the odd multipoles at small $l$ ($l<20$). The suppression of even multipoles was studied by  \cite{Land:2005jq} for the first-year maps of WMAP and it was shown that the \emph{sawtooth} pattern of the angular power spectrum at small $l$ was not statistically significant although visually striking.

As opposed to the full-sky maps, masked maps have a small probability: only $0.73\%$  of simulations had an $A(x)$ smaller than that of ILC map masked with the kp0 intensity mask. Q and W frequency band maps have the same probability and for the V band it is $0.69\%$. For the TOH map this number is $0.95\%$. 
This quantifies the smallness of the two-point correlation on large scales of cut-sky maps (see Table \ref{tab:table1}). These results qualitatively agree with the results of \cite{Copi:2006tu}.
\begin{table}
\caption{\label{tab:table1} The significance of the smallness of
$C(\theta)$ on large angular scales for
different sets of maps and masks discussed in this paper. The third 
column is the fraction of 10,000 simulations that has an $A(x)$ 
smaller than the $A(x)$ derived from the 3-year data of WMAP. This shows that \emph{a)} cut-sky maps are discrepant on large scales while full-sky maps are not, \emph{b)} masking the Galactic center and the region around the Gum Nebula in full sky CMB maps (8.9\% of the sky), \texttt{mask1}, has almost the same effect as the kp0 mask (23.5\% of the sky), while applying the complimentary mask (the rest of the kp0 mask with these two spots excluded, see Fig. \ref{fig:mask1}) does not suppress the large scale power much, and \emph{c)} the quadrupole and the octupole happen to have the magnitude and orientation needed for making the two-point correlation in  cut-sky maps very small on large angular scales. Excluding them from the maps or changing their orientation brings the large scale power back.}
\begin{ruledtabular}
\begin{tabular}{lcr}
map&mask&$P(A_{sim}<A_{data})$\footnote{$A(x)$ is calculated at $x\simeq 0.53$ (the minimum of the $A(x)$ for cut-sky maps). }\\
\hline
ILC &(full sky) & 8.1\% \\
TOH&(full sky) & 7.4\% \\
\hline
Q &(kp0)& 0.73\%\\
V &(kp0) &  0.69\% \\
W &(kp0)& 0.73\% \\
ILC& (kp0)&  0.73\%\\
TOH &(kp0) &   0.95\%\\
\hline
ILC ($C_2=0$)& (kp0) & 7\%\\
ILC ($C_2=C_3=0$)& (kp0) & 40\%\\
ILC (rotated $l$=2)& (kp0) & 6.50\%\\
ILC&(mask1)& 0.78\% \\
ILC&(kp0-mask1)&  12\%\\

\end{tabular}
\end{ruledtabular}
\end{table}

Is the low power in the cut-sky maps due to the low quadrupole? 
Removing the quadrupole changes the shape of the two-point correlation to some extent. But removing $l=2$ and $l=3$ multipoles together seems to have a bigger effect on the shape of the $C(\theta)$ (Fig. \ref{fig:multipoles}). Interestingly, the contribution from the $l=2,3$ multipoles to the two-point correlation on large scales is almost equal to the contribution of the rest of multipoles but with a negative sign. That is 
\begin{equation}
  \label{eq:3}
  \sum_{l=2,3} (2l+1) C_l P_l(\cos\theta)|_{\theta>60^\circ} \simeq -  \sum_{l>3}  (2l+1) C_l P_l(\cos\theta)|_{\theta>60^\circ}.
\end{equation}
To quantify this effect, we compute the $A(x)$ for quadrupole and octupole removed maps. The simulations are also made from the best fit LCDM power spectrum with $C_2=C_3=0$, and  $40\%$ of simulations have an $A(1/2)$ smaller than that of the cut-sky ILC map (see Table \ref{tab:table1}), which confirms the effect of the quadrupole and octupole on making the large scale power small, and indicates that the magnitudes of the low multipoles happen to be such that they cancel the contribution from the rest of the multipoles to the correlation function on large scales.

 The fact that smallness of two-point correlation on large scales happens in the cut-sky maps suggests that the orientation of the low multipoles relative to the Galactic plane may also play a role. To examine that, we keep the magnitude of the quadrupole fixed and change its orientation. This can be done by permuting the harmonic coefficients, $a_{2m}$, of the full-sky ILC map in the following way\footnote{This is not a general way of rotating the $a_{lm}$ coefficients, but is an example of a possible way of changing the orientation of a multipole by preserving the $C_l$. A general way of doing so is by the means of Wigner rotation matrices (see \it{e.g.} \cite{Hajian:2005jh}).}:
$$a_{20} \rightarrow a_{20}, \,\,\, a_{21} \rightarrow a_{22} \rightarrow  a_{21}.$$
A CMB anisotropy map constructed with this new quadrupole will have the same statistical properties as the ILC map (including the same power spectrum), but the orientation of the quadrupole in this map is different from that of the ILC map.  We apply the kp0 mask to this map and compute the two-point correlation, which appears to be normal.  The $A(1/2)$ of this map is improved: the fraction of simulations that have smaller $A(1/2)$ than this map increases from $0.73\%$ to $6.5\%$. This proves that not only the 
magnitude of the low multipoles but also their orientation is responsible for making the large scale power of the cut-sky maps negligible.

The fact that  the large scale power of the WMAP data comes from the galactic plane is strange enough, but we can hunt it down further to see what parts of the Galactic plane contain most of the power. To do so we break the kp0 mask into different sections to make new masks. $A(1/2)$ is then computed for the ILC map masked with each of these new masks and compared to 10,000 simulations of CMB anisotropy sky masked with the same masks. We see that most of the large scale power comes from two spots, a cold spot at $ -41^\circ < \theta < 37^\circ$ and  a hot spot  at $ 244^\circ <\theta <276^\circ $, both regions within the kp0 intensity mask.  We make a new mask to cover those two spots (\texttt{mask1}). This mask excludes 8.9\% of the pixels from the maps. $C(\theta)$ computed from the ILC map masked with \texttt{mask1} is flat and close to zero on large angular scales and has a shape  similar to the $C(\theta)$ derived from the ILC map masked with kp0. Also $A(1/2)$ is  small for this map and $P(A_{sim}<A_{mask1})=0.78\%$. On the other hand the complementary mask to\texttt{ mask1}, which is the rest of the kp0 mask, has a  small effect on the shape of $C(\theta)$ on large angular scales and $C(\theta)$ for the ILC map masked with this complementary mask is  close to $C(\theta)$ of the full-sky ILC map on large scales. Also $A(1/2)$ is not small for the ILC map masked with this complementary mask and  $P(A_{sim}<A_{kp0-mask1})=12\%$ (see Table \ref{tab:table1}). This mask has a small effect on the octupole but changes the orientation of the quadrupole. This is in agreement with the observation of \cite{Bielewicz:2004en} that  the quadrupole (unlike the octupole) has a non-negligible dependence on the applied mask and foreground correction. 

\begin{figure}
 \includegraphics[width=0.3\textwidth, angle=90]{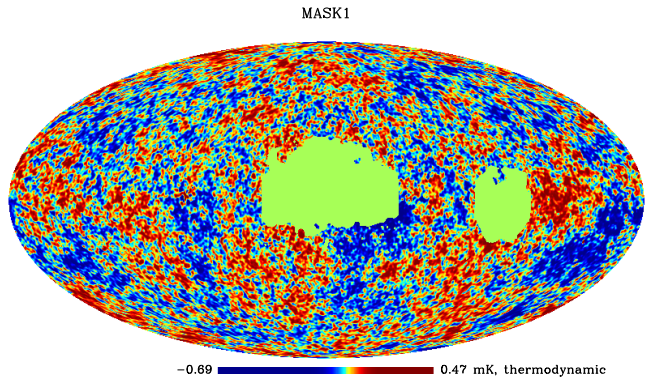}
 \caption{\label{fig:mask1} Masking these two spots ($\sim 9\%$ of the sky) in the ILC map makes the two-point correlation of the resultant map small on angular scales larger than 60 degrees.}
 \end{figure}

Our analysis shows that most of the large scale power of CMB anisotropy maps comes from two regions around the Galactic plane. One of them is a cold spot within $\sim 40^\circ$ of the Galactic center and the other one is a hot spot in the vicinity of the Gum Nebula. Although these are the potential suspects for foreground contamination \cite{Finkbeiner:2003im,deOliveira-Costa:2006zj}, the discrepancy arises when one masks them out. 

It might be possible that the actual power of the CMB anisotropy is intrinsically low on large angular scales. There are a few theoretical models that predict suppression of power on large scales. Spatially compact spaces are examples of these models. Although compact spaces with  fine tuned matter (and total) densities can explain the suppressed large scale power of the WMAP data (see \emph{e.g.} \cite{Gundermann:2005hz, mythesis}), they fail  to explain quadrupole-octupole alignment \cite{Weeks:2006rr}, are not found by the circles-in-the-sky search \cite{ShapiroKey:2006hm} and are inconsistent with the bipolar power spectrum analysis of the WMAP data \cite{mythesis, Hajian:2006ud, Hajian:2005jh}. Other possibilities such as fine tuned Bianchi models are already ruled out because their model parameters are not consistent with the observed cosmological parameters and particularly because their total density is too low (see \cite{Ghosh:2006xa} for a complete review on these models). Other possibilities such as homogeneous local dust-filled voids that can suppress the fluctuations due to the linear ISW effect have recently been studied  \cite{Inoue:2006fn}.

At present more data are needed to determine the nature of the regions that contain most of the large scale power. If the current foreground modeling is correct then the only remaining peculiarity is the $l=2$ and $l=3$ alignment \cite{Copi:2005ff}. If the current foreground modeling is incorrect and there is no CMB power on large angular scales, then we will have to find a better cosmological model or live with a $\simeq 1\%$ (a posteriori) chance for our current preferred model.

\acknowledgements
I would like to thank Lyman Page and David Spergel for 
encouraging me to work on this problem and for enlightening
discussions throughout this work. I am thankful to Toby Marriage, Jeff Weeks, Dragan Huterer, Olivier Dore, Cristian Armendariz-Picon, Licia Verde and Kate Land for their comments on the manuscript that helped 
me improve the paper.
Some of the results in this paper have been derived using 
the HEALPix package \cite{Gorski:2004by}. 
I acknowledge the use
of the Legacy Archive for Microwave Background Data Analysis
(LAMBDA). Support for LAMBDA
is provided by the NASA Office of Space Science.
Support for this work was provided by NASA grant LTSA03-0000-0090.

\end{document}